\documentclass{aastex}          
\usepackage{spr-astr-addons}    

\usepackage{graphicx}	
\usepackage{amsmath}	
\usepackage{amssymb}	
\usepackage{threeparttable}
 \usepackage{booktabs}
\usepackage{longtable,booktabs}
 \usepackage{multirow}
 \usepackage{color}

\newcommand{\emaila}{}

\begin{document}

\title{Timing irregularities  of PSR~J1705$-$1906}

\shorttitle{<Short article title>}
\shortauthors{<Autors et al.>}

\author{Y. L. Liu\altaffilmark{1,2}}
\and
\author{J. P. Yuan\altaffilmark{$^\dagger$,2,3,4}}
 \and
\author{J. B. Wang\altaffilmark{2,4}}
\and
\author{X. W. Liu\altaffilmark{1}}
\and
\author{N. Wang\altaffilmark{2,3,4}}
\author{R. Yuen\altaffilmark{2,3}}

\altaffiltext{$^\dagger$}{\emaila{E-mail: yuanjp@xao.ac.cn}}
\altaffiltext{1}{School of Physics, China West Normal University, Nanchong,
Sichuan, China, 637002}
\altaffiltext{2}{Xinjiang Astronomical Observatory, CAS, 150 Science 1-Street,
 Urumqi, Xinjiang, China, 830011}
\altaffiltext{3}{Key Laboratory of Radio Astronomy, Chinese Academy of
Sciences, Nanjing 210008, China}
\altaffiltext{4}{Xinjiang Key Laboratory of Radio Astrophysics, 150
Science 1-Street, Urumqi, Xinjiang, 830011, China}

\begin{abstract}\\
 Timing analysis of PSR J1705$-$1906  using data from Nanshan
 25-m and Parkes 64-m radio telescopes, which span over fourteen years, shows that  the pulsar exhibits significant proper motion, and rotation
instability. We updated the astrometry parameters and the spin parameters  of the pulsar.
In order to minimize the effect of timing  irregularities on measuring  its
position, we employ the Cholesky method to analyse the timing noise.
 We obtain the proper motion of $-$77(3) \,mas\,yr$^{-1}$ in right ascension and $-$38(29) \,mas\,yr$^{-1}$ in declination.
The power spectrum of timing noise is analyzed for the first time,  which gives the
spectral exponent $\alpha=-5.2$ for the power-law  model indicating that the fluctuations  in spin frequency and spin-down rate
 dominate the red noise.
 We detect two small glitches  from this pulsar with fractional jump  in spin
frequency of $\Delta \nu/\nu\sim2.9\times10^{-10}$ around MJD~55199 and
$\Delta \nu/\nu\sim2.7\times10^{-10}$ around MJD~55953.
Investigations of pulse profile at different time segments suggest no significant changes in the
pulse profiles around the two glitches.
\end{abstract}


\section{Introduction}\label{s:Introduction}
Discovered in the second Molonglo pulsar survey \citep{manchester1978}, PSR~J1705$-$1906 is a radio and $\gamma$-ray pulsar with a  rotation period of 0.299\,s and period derivative of $4.1\times10^{-15}\rm{\,s\,s^{-1}}$.
 The pulsar has a characteristic age  $\tau_{\rm c} \sim 1.16\times10^6$ yr, and  a spin-down energy loss rate of $6.11\times 10^{33}\,\rm{erg\,s^{-1}}$.
Assuming magnetic dipolar braking, PSR J1705$-$1906 has a  surface dipole magnetic field $B_{\rm s}\sim 1.12\times10^{12}$\,G.
 High-energy pulsed emission  was also detected from the pulsar using the Fermi Gamma-ray telescope, which revealed a $\gamma$ ray luminosity of $0.25(8)\times10^{33}$ erg $\rm{s^{-1}}$  with efficiency of 4\%,  and a spectral index  of $2.3(2)$ \citep{hou2014+1}.
The pulse profile exhibits a clear single peak between 0.1 - 300\,GeV, which can be well-fitted with a Lorentz function.
Note that the Fermi telescope has no detection of cut-off energy.  X-ray counterpart  
 was not confirmed with the XMM-Newton 8\,ks observation  between 0.3  and several\, keV bands \citep{hou2014+1}.

It has long been recognized that pulsars are high-velocity objects with space velocities up to an order of magnitude larger than  that of their progenitors \citep{gunn1970,hobbs2005}.
A natural explanation for such high velocities is that supernova explosions are  non-homogeneous resulting in kicks to the  nascent pulsars \citep{lai1998,lai2001,ng2006}.
Pulsar velocities are determined by measuring their proper motions and distances.
For  radio pulsars with high intensity, their positions and proper motions can be precisely determined  using Very Long Baseline Interferometry (VLBI) \citep{brisken2002,brisken2003,deller2016}.
In recent years, the proper motion of some weak pulsars have also been obtained by VLBI.  For example, \citet{yan2013} measured the astrometric parameters of PSR~B1257+12 in the  three-planet pulsar  system using Very Long Baseline Array (VLBA) and European VLBI Network (EVN), and \citet{du2014} acquired the parallax and proper motion for the weak millisecond pulsar PSR~J0218+4232 with EVN at 1.6\,GHz. In addition, astrometric parameters can be determined  by pulsar timing observations  with data that  spans over several years \citep{manchester1974,hobbs2005,li2016}.
PSR~J1705$-$1906 is a nearby pulsar with  an estimated distance of 0.9(1)\,kpc based on the Galactic free electrons density model (NE2001), but its parallax is not reported.
Previous studies of PSR~J1705$-$1906 reported its positions close to the ecliptic plane with ecliptic longitude of $257.135499^{\circ}$, ecliptic latitude of $3.72327^{\circ}$, right ascension of 17:05:36.099 and declination of $-$19:06:38.6 \citep{hobbs2004}.
The proper motion of the pulsar in ecliptic coordinates is presented with $\mu_{\lambda}=-66(5)$\,mas\,yr$^{-1}$ and $\mu_{\beta} =-123(83)$\,mas\,yr$^{-1}$, as well as in equatorial with $\mu_{\alpha}=-78(9)$\,mas\,yr$^{-1}$ and $\mu_{\delta}=-116(82)$\,mas\,yr$^{-1}$\citep{hobbs2004}.

Pulsar timing is a high-precision discipline, which allows the observed pulse Time of Arrivals (ToAs) to be compared with a model of the pulsar's astrometric, orbital and rotational, parameters \citep{hobbs2006}.
The difference between the predicted arrival time and the actual arrival time is known as the pulsar's timing residual.
Long-term timing observations reveal two main sources of timing irregularities, namely the timing noise and glitch, where the latter exhibits as a sudden speed-up in the spin-down rate of a pulsar.
Timing noise is a continuous behavior with low frequency quasi-sine structure (called red noise) over time scale of months to years, which is commonly caused by irregularities in the intrinsic pulsar rotation.
For most observed glitches, the increase in spin-frequency, $\nu$, ranges from $10^{-9}$\,Hz to $10^{-5}$\,Hz \citep{yuan2010,espinoza2011,yu2013}.
For PSR~J1705$-$1906, a glitch with $\Delta\nu/\nu\sim0.4\times10^{-9}$ was detected in 1992 \citep{espinoza2011}.
After a glitch, there are often exponential recovery in spin-frequency ($\nu$) and decay of the frequency derivative ($\dot{\nu}$) back to their pre-glitch values. This relaxation can vary significantly from pulsar to pulsar, and even from glitch to glitch in the same pulsar \citep{alessandro1996}.
A more complete understanding of pulsar timing irregularities will lead to many important results, such as explaining the cause of timing noise and glitches, which may also allow correlations to form between these phenomena and hence provide unique insights into the interior structure of neutron stars.

This paper is organized as follows. We describe the observing system and the data reduction of PSR J1705$-$1906 in Section 2.
In Section 3, we obtain the latest information on astrometric measurements for the pulsar.
In Section 4, we present the analysis of red timing noise and describe the two glitches.
In Section 5,  we investigate whether there is any change in the pulse profile  of PSR J1705$-$1906 associated these glitches.
We discuss our results in Section 6.
\section{Observations and data reduction}\label{s:Observations and data reduction}
\begin{table*}
\caption{ Description of the observations showing the telescope, receiver, centre frequency and the bandwidth (BW). The last two columns are the number of ToAs and the MJD ranges over which the observations spaned.}
\label{tbl:dataset}
\begin{tabular}{llcrrrc}

\toprule[1px]
Telescope & Receiver     &    Freq.(MHz) & BW(MHz)  & Backend & No.of ToAs & MJD range\\
\hline
Nanshan   & (non-)cryogenic   & 1540 &  320 $~$$~$ &   AFB   & 510 $~$ $~$         & 51561--56717 \\
Nanshan   & cryogenic         & 1556 &  320 $~$$~$ & PDFB3   & 120 $~$ $~$         & 55239--56720 \\
Parkes    & MULTI             & 1374 &  288 $~$$~$ &   AFB   &  12 $~$ $~$         & 52001--54009 \\
Parkes    & MULTI             & 1369 &  256 $~$$~$ & PDFBs   &  86 $~$ $~$         & 54303--56740 \\
Parkes    & 1050CM            & 3094 & 1024 $~$$~$ & PDFBs   &  16 $~$ $~$         & 54305--56683 \\
\bottomrule[1px]
\end{tabular}
\end{table*}

The Nanshan telescope is equipped with a dual-channel cryogenic receiver operating at frequency band centered at 1.54\, GHz with a bandwidth of 0.32\,GHz.
Early observations, prior to 2010, were taken using an Analogue Filter-Bank (AFB) that has 128 2.5\,MHz sub-channels for each of the two polarizations \citep{wang2001}.
Since January 2010, observations have been made using the Digital Filter-Bank (DFB) with 0.5 MHz bandwidth for each sub-channel where four Stokes parameters are collected from the two orthogonal polarizations.
Data is folded on-line with sub-integration time of 1\,min for the AFB and 30\,s for the DFB, and then written to disk with 256\,bins across the pulse profile for the AFB and 512\,bins for the DFB.
PSR~J1705$-$1906 was observed roughly three times per month with 4 minutes each time at Nanshan.
The observations at Parkes were carried out between April 2001 and Feb 2014 with a central observing frequency close to 1.37\,GHz. Several observations of this pulsar were conducted with the 10-cm receiver which has a bandwidth of 1024\,MHz centered at 3094\,MHz.  The raw data is acquired by using a suite of Parkes digital filter-bank systems \citep{hobbs2011} and all of the pulsar data was stored in the data archive\footnote{\url{https://datanet.csiro.au/dap}} following the $\textsc{psrfits}$ software \citep{hotan2004}. The details  for receivers, back-ends and the  data set can be found in Table \ref{tbl:dataset}.

The $\textsc{psrchive}$ analysis system is used to dispel radio-frequency interference, fold and de-disperse the multi-channel data to get effective pulse profiles \citep{hotan2004}. The $\textsc{tempo2}$ software is then used to calculate the barycentric arrival time, form the timing residuals and carry out the weighted least-square fit \citep{hobbs2006,edwards2006}. The timing model for barycentric pulse phase, $\phi$, as a function of time
$\emph{t}$, is,
\begin{equation}
\label{eq:phase}
 \phi(t) =\phi_{0}+\nu(t-t_{0})+\frac{1}{2}\dot{\nu }(t-t_{0})^{2}+  \frac{1}{6}\ddot{\nu } (t-t_{0})^3,
\end{equation}
where $\phi_{0}$ is the phase at time $t_{0}$, and $\nu$, $\dot{\nu}$, $\ddot{\nu}$ represent the pulse frequency, frequency derivative and
 frequency second derivative, respectively. A glitch will result in an additional pulse phase that can be modeled by the equation:
\begin{equation}
\label{eq:glitch}
\begin{split}
\phi _{\rm{g}}=\Delta \phi +\Delta \nu _{\rm{p}}(t-t_{\rm{g}})+\frac{1}{2}\Delta \dot{\nu } _{\rm{p}}(t-t_{\rm{g}})^{2} \\
+[1-e^{-(t-t_{\rm{g}})/\tau _{\rm{d}}}]\Delta\nu_{\rm{d}}\tau_{\rm{d}},
\end{split}
\end{equation}
where $t_{\rm g}$ is the glitch epoch and $\Delta\phi$ is an offset of pulse phase between the pre- and post-glitch data. The glitch event is
characteristic with permanent increments in the spin frequency $\Delta\nu_{\rm p}$ and first frequency derivative $\Delta\dot{\nu}_{\rm p}$ and a transient
frequency increment  $\Delta\nu_{\rm d}$ which decays exponentially with a time scale $\tau_{\rm d}$.

\begin{figure}[t]
\begin{center}
\centerline{\includegraphics[angle=0,width=0.45\textwidth]{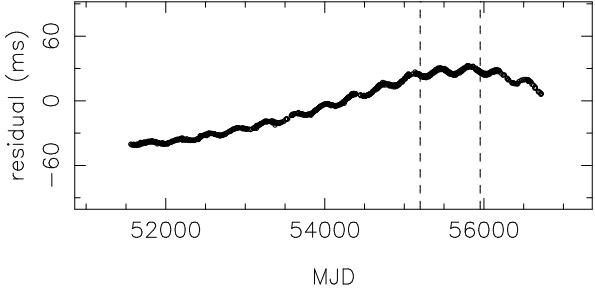}}
\caption{The timing residuals for PSR J1705$-1906$ with respect to the ephemeris given by \citet{hobbs2004}.}
\label{fig:allresidual}
\end{center}
\end{figure}

\begin{table}
\caption{ The ephemeris for PSR J1705$-$1906 \citep{hobbs2004}. }
\label{tbl:premodel}
\begin{tabular}{ll}
\toprule[1px]
$\rm{NAME}$                    &   PSR\,J1705$-$1906\\
$\rm{RAJ}$  (h:m:s)            &  17:05:36.099(3)\\
$\rm{DECJ}$ ($^{\circ}:':''$)  &  $-$19:06:38.6(4)\\
$\rm{DM}$ (pc cm$^{-3}$)                   &   22.907(3)\\
$\rm{PEPOCH}$ (MJD)            &   48733.00\\
$\rm{POSEPOCH}$ (MJD)          &   48733.00\\
$\rm\nu$ (Hz)                 &     3.344622243443(18)\\
$\rm\dot\nu$ ($\times 10^{-15}~$Hz s$ ^{-1}$)        &    $-$46.28835(11) \\
$\rm\ddot\nu$ ($\times 10^{-15}~$ Hz s $^{-2}$)       &   0.334(15)\\
$\rm{EPHVER}$              &    2\\
$\rm{UNITS}$                 &     TDB\\
\bottomrule[1px]
\end{tabular}
\end{table}

\begin{table}
\caption{The  positions of PSR~J1705$-$1906  in equatorial coordinate determined at the center epoch for each of the three date segments. The values in parentheses indicate uncertainties in the unit of the last quoted digit.}
\scriptsize
\label{tbl:pos1}
\begin{tabular}{llll}

\toprule[1px]
$\rm{Data}$ $\rm{span}$& $\rm{Epoch}$     & $\rm{RA}$        & $\rm{DEC}$      \\
(MJD)             &          (MJD)     &      (h:m:s)       & ($^{\circ}:':''$)    \\
\hline
51560-55199     & 53361                 & 17:05:36.031(3)  & $-$19:06:39.3(3)  \\
55199-55953     & 55573                 & 17:05:35.9969(14)& $-$19:06:39.51(16) \\
55953-56741     & 56350                 & 17:05:35.9851(18)& $-$19:06:39.42(12) \\
\bottomrule[1px]
\end{tabular}
\end{table}

\begin{figure}
\centerline{\includegraphics[angle=0,width=0.48\textwidth]{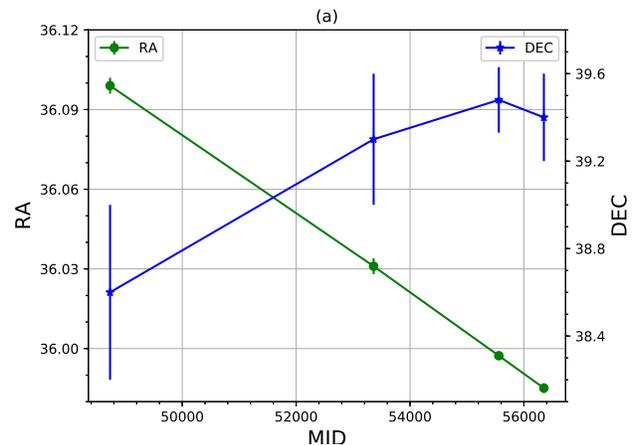}}
\vspace{-2mm}
\caption{Change of the right ascension and declination along with epoch.}
\label{fig:position1}
\end{figure}

\section{Astrometry of PSR~J1705$-$1906}\label{s:Astrometry}
\subsection{Position and proper motion}\label{ss:Position and proper motion}
 Figure~\ref{fig:allresidual} shows the timing residuals for PSR~J1705
$-$1906, which is obtained using the ephemeris (shown in Table~\ref{tbl:premodel})
from the ATNF Pulsar Catalogue\footnote{\url{http://www.atnf.csiro.au/research/pulsar/psrcat/}}.
The residuals appear to exhibit periodic oscillation on a time scale of one year owing to
the proper motion.
As the two glitches occurred around MJD~55199 and MJD~55953 (detail are given in subsequent analysis), the timing data is divided into three segments to obtain position and proper motion more accurately.
 The two glitches are not obvious in Figure ~\ref{fig:allresidual} as they are submerged in the
residuals due to the proper motion.

We analyze the position and proper motion of PSR 1705$-$1906 with the three data segments. The first segment (between MJD~51560 to MJD~55199) has significant red noise after fitting for the position, spin frequency, frequency derivative and second derivative of the frequency. It is always a challenge to whiten the timing residuals when the data is dominated by red noise, which is a well-known problem to pulsar observers. To deal with it usually involve subtracting a high order polynomial or harmonically related sinusoids when carrying out the least squares fit. \citet{coles2011} have shown that such technique is not optimal and often leads to a severe underestimation of the parameter error, and so proposed a prewhitening method known as the Cholesky method.
The Cholesky solution uses a linear transformation, which is derived from the covariance matrix of the residuals, to pre-whiten  the residuals. It estimates an analytic model for the low frequency component of the spectrum for the timing noise (see Section \ref{ss:Timing noise} for details).
We fit the pulsar parameters (including the position and proper motion) with the analytic red noise model, and obtained the right ascension $\alpha$=17:05:36.031(3) and declination $\delta=-$19:06:39.3(3) at MJD~53361 in equatorial coordinates as shown in Table~\ref{tbl:pos1}.
These values
will be adopted in the subsequent analysis. 
As the data in the last two segments are not found to have non-white noise after fitting the position, $\nu$, $\dot{\nu}$, and $\ddot{\nu}$, we employ the polynomial fitting for the high-order derivatives of the spin frequency, not the Cholesky method, to estimate the positions.
Table~\ref{tbl:pos1} shows positions at the center epoch for each of three segments in equatorial coordinates.

We plot the positions of the pulsar in the four epochs in Figure~\ref{fig:position1} to better show the changes of  position. 
The first position in MJD 48733 (21 April 1992) is obtained from  the paper by \citet{hobbs2004}.
In this work, we found that the uncertainties for the position of PSR~J1705$-$1906 in declination is greater than that in  right ascension.
The reason is that the pulsar is subjected to strong roemer delay and shapiro delay if it lies in or very close to the ecliptic plane.
Therefore, the predicted positions and proper motions in ecliptic latitude are relatively poor, nor can we precisely measure its position and proper motion in declination, as the angle between the equator and the ecliptic is small of $\sim$23.5 degrees.

Proper motion of the pulsar is calculated using
\begin{align}
\label{eq:proper1}
\mu_{\alpha}=\dot\alpha cos \delta ~~~\mu_{\delta}=\dot\delta,
\end{align}
\begin{align}
\label{eq:proper3}
\mu_{tot}=(\mu_{\alpha}^{2}+\mu_{\delta}^{2})^{\frac{1}{2}},
\end{align}
where $\mu_{\alpha}$ is proper motion in right ascension, $\mu_{\delta}$ is proper motion in declination and $\mu_{tot}$ is total proper motion in two dimensions.
We derive the proper motion using the positions obtained from the first segment and \citet{hobbs2004}. 
The results are 
$\mu_{\alpha}=-76(5)$ mas\,yr$^{-1}$ and $\mu_{\delta}=-55(39)$\,mas yr$^{-1}$, which are consistent with the measurements of $\mu_{\alpha}=-78(9)$\,mas\,yr$^{-1}$
and $\mu_{\delta}=-116(82)$\,mas\,yr$^{-1}$ given by \citet{hobbs2004}. Our results have smaller uncertainties as they are derived from a longer data span.

\subsection{Distance and velocity}\label{ss:Distance and velocity}
Distance and transverse velocity of the pulsar determine the luminosity of the source and its location in the Galaxy, 
all of which are important for research of the origin, evolution and emission properties of the object in question.
The distribution of free electrons in the Galaxy, the Magellanic Clouds and the intergalactic medium can be used to estimate distance to a pulsar based on its dispersion measure.
We calculated the distance to PSR J1705--1906 using the latest electronic density model $\textsc{ymw16}$ \citep{ymw16} and the dispersion measurement (DM) of $22.907(3)\,\rm{pc\,cm^{-3}}$ \citep{hobbs2004}.
A distance of 0.7469\,kpc is obtained from $\textsc{ymw16}$, which is smaller than 0.89 kpc predicted by NE2001.
We note that 68\% of the distances predicted by $\textsc{ymw16}$ will have a relative uncertainty of less than 0.4.
The pulsar transverse velocity is determined with $V_{T}=4.74\mu_{tot}D\approx304(57)~$km s$^{-1}$,
where $D$ is distance (in kpc) and the uncertainty in $\mu_{tot}$ is dominated by the error in proper motion and distance. \citet{lyne1994} have demonstrated that the  average transverse velocity of pulsars is  $V_{T}=\rm{300(30)\,km\cdot s^{-1}}$ based on analysis of proper motions and distances from a large number of pulsars.
It seems that PSR~J1705$-$1906 has a medial transverse velocity among pulsars.

\section{Timing irregularities}\label{s:Timing irregularities}
\subsection{Timing noise}\label{ss:Timing noise}

\begin{table}
\scriptsize
\caption{The $\Delta_8$ values of PSR~J 1705$-$1906.}
\label{tbl:noise8}
\begin{center}
\begin{tabular}{cccc}
\toprule[1px]
 Data span     &   $\nu$                & $\ddot{\nu}$                           & $\Delta_{8}$     \\
          (MJD)   &  ($\rm{s^{-1}}$)       &($10^{-26}\,\rm{s^{-3}}$)           &                              \\
 \hline
 51560-52764 &      3.344608546886(8)    &12(3)                           & $-$2.17(11)               \\
 52764-53968 &     3.344603712604(5)    &4(2)                            & $-$2.65(22)             \\
 53968-55199 &      3.344598894692(4)    &3(1)                            & $-$2.77(14)               \\
\toprule[1px]
\end{tabular}
\end{center}
\end{table}

\begin{figure}
\centerline{\includegraphics[angle=0,width=0.45\textwidth]{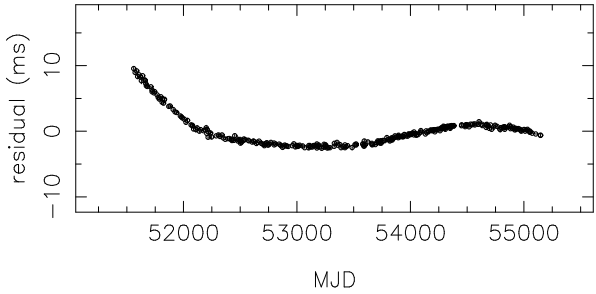}}
\vspace{-2mm}
\caption{The timing  residuals of PSR~J1705$-$1906 between MJD~51560  and MJD~55199.}
\label{fig:residual}
\end{figure}
\begin{figure}
\centerline{\includegraphics[angle=0,width=0.45\textwidth]{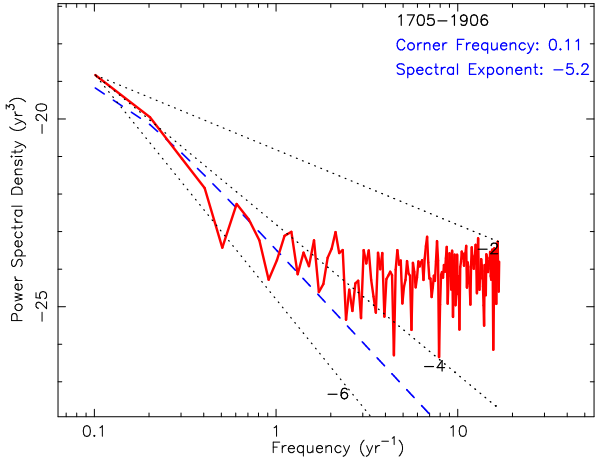}}
\vspace{-2mm}
\caption{The observed power spectra of the pre-glitch timing noise for PSR~J1705$-$1906. The red solid line is the observed spectral noise, the blue dashed line represents the red noise model. The dotted lines show power spectra with exponent of $-$2, $-$4 and $-$6 from top to bottom, respectively.}
\label{fig:spectral}
\end{figure}

Figure~\ref{fig:residual} shows the pre-glitch timing residuals with respect to a solution that contains the $\nu$ and $\dot{\nu}$ for PSR~J1705$-$1906. It is clear that pre-glitch timing residuals exhibit significant red noise. In order to characterize the stability of the pulsar and to determine the ``amount of timing noise'', we calculate the $\Delta_8$ value \citep{arzoumanian1994} given by
\begin{equation}
\label{eq:rednoise}
\Delta_{8}=log(\frac{\lvert\ddot\nu\lvert}{6\nu }t^{3}),
\end{equation}
where $\nu$ and $\ddot{\nu}$ are measured over a time interval of $t\sim10^{8}\,\rm{s}$ ($\sim$3 yr).
As this pulsar's pre-glitch data-sets span at least 10\,yr, we obtained the $\Delta_8$ values of $ -2.17(11),-2.65(22),-2.77(14)$ by fitting  $\nu$, $\dot{\nu}$ and $\ddot{\nu}$ to three $\sim3$ yr  segments.
 \citet{hobbs2010} confirmed the  statistics-averaged correlation between $\Delta_8$ and spin-down rate, $\rm{\dot{P}}$,  for 366 non-recycled pulsars
\begin{equation}
\label{eq:rednoise1}
{\Delta_8}' = 5.1 +0.5log\rm{\dot{P}}.
\end{equation}
 With the observed $\rm{\dot{P}}$, the inferred value ${\Delta_8}'$ of this pulsar is $\sim -2.09$.
Table~\ref{tbl:noise8} presents the data spans, epochs, the values of  $\nu$, $\ddot{\nu}$ 
and $\Delta_8$ for the three data segments. 
 It is obvious that one of values for $\Delta_8$ ($-$2.17) is close to the correlation, although the measured ones are scattered.

We use the Cholesky method for the analysis of correlated timing noise. A power-law model, given by
\begin{equation}
 P(f)=A[1+(f/f_c)^2]^{\alpha/2},
\end{equation}
can be used to fit the low-frequency noise. Here, $\emph{A}$ is the amplitude, $f_c$ is the corner frequency and $\alpha$ is the spectral exponent.
We estimate $f_c$ and $\alpha$ to match the red spectrum at low frequencies in $\textsc{tempo2}$.
For the pre-glitch data, these values are $f_{c}=0.11\,\rm yr^{-1}$ and $\alpha=-5.2$ for a reasonable fit.
Figure~\ref{fig:spectral} shows the distribution of power spectra density against frequency. The dotted lines embody the spectral exponent of $-2$, $-4$, and $-6$, which suggest that the noise is dominated by random walk in the phase, $\nu$ and $\dot\nu$, respectively. The blue dashed line represent the low-frequency model with spectral exponent of $\sim -5.2$, indicating that the fluctuations  in spin frequency and spin-down rate dominate
the red noise.
Rotation parameters of
$\nu=3.344603736538(22)\,\rm{s^{-1}}$,
$\dot\nu=-4.62782(2)\times10^{-14}\,\rm{s^{-2}}$ and
$\ddot\nu=4.7(7)\times10^{-26}\,\rm{s^{-3}}$
are obtained at MJD 53361 by Cholesky method, which are adopted in the subsequent analysis.

\subsection{Glitch}\label{ss:Glitch}

\begin{figure}
\centerline{\includegraphics[angle=0,width=0.44\textwidth]{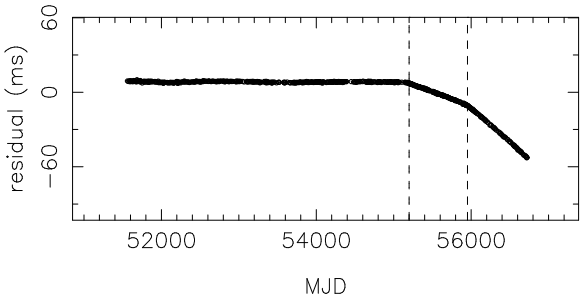}}
\vspace{-2mm}
\caption{The timing residuals of PSR~J1705$-$1906 after updating the latest astrometry parameters and spin parameters.}
\label{fig:resiglt}
\end{figure}

\begin{figure}
\centerline{\includegraphics[angle=0,width=0.44\textwidth]{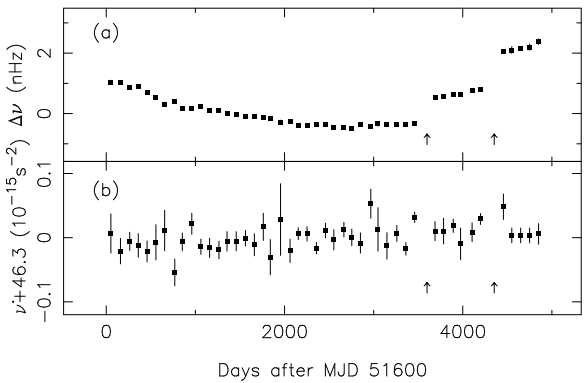}}
\vspace{-2mm}
\caption{Variations of the spin-frequency residuals $\nu$ showing two glitches that occurred around MJD~55199 and MJD~55953}
\label{fig:glt}
\end{figure}

\begin{table*}
\caption{Rotation parameters of PSR~J1705$-$1906 obtained from different segments.}
\label{tbl:slt}
\begin{center}
\begin{tabular}{lllllll}
\toprule[1px]
 Data sapn &Epoch & $\nu$          & $\dot\nu$       & $\ddot\nu$   &  No. of TOAs & post-fit Rms \\
 &(MJD)   & ($\rm s^{-1}$)   & ($\times10^{-14}~\rm s^{-2}$)    & ($\times10^{-26}~\rm s^{-3}$) &    &  ($\mu s$)     \\
\hline
51561 - 55199 & 53361  & 3.344603736531(1)  &  $-4.627826(2)$   &  $5.01(5)$           &    344   & 115.994\\
55199 - 55953 & 55573  & 3.344594893724(7)  &  $-4.62664(3) $   &  $1.9(6)$            &    130   & 70.102 \\
55953 - 56741 & 56350  & 3.344591788845(5)  &  $-4.62611(5) $   &  $1.9(8)$            &    95    & 168.725\\
\toprule[1px]
\end{tabular}
\end{center}
\end{table*}

\begin{table*}
\caption{ Glitch parameters for PSR~J1705$-$1906.}
\label{tab:glitch}
\begin{center}
\begin{tabular}{llllllll}
\toprule[1px]
               &       & \multicolumn{2}{c}{Extrapolated}    &  \multicolumn{2}{c}{Fitted}   &     &              \\
 Data span  &  Epoch  &  $\Delta\nu_{\rm{g}}/\nu$  &  $\Delta\dot{\nu}_{\rm{g}}/\dot{\nu}$  &  $\Delta\nu_{\rm{g}}/\nu$  &  $\Delta\dot{\nu}_{\rm{g}}/\dot{\nu}$ & No. of TOAs  &  post-fit Rms \\
               & (MJD)    & ($\times10^{-9}$) & ($\times10^{-3}$)  & ($\times10^{-9}$) & ($\times10^{-3}$)  &     & ($\mu s$)    \\
\hline
51560$-$55372  & 55199  & 0.26(1)  & 0.4(4)                       & 0.29(5)  & 0.3(5)                       & 378 &  140.156     \\
55199$-$56153  & 55953  & 0.28(2)  & 0.2(1)                        & 0.27(1) &  0.5(2)                      & 159 &  68.512      \\
\toprule[1px]
    \end{tabular}
  \end{center}
\end{table*}

We present the analysis of glitches using Nanshan and Parkes timing data that was collected from 2000 January to 2014 March.
The latest astrometric parameters and red noise model are fixed in this analysis. The timing residuals are obtained by comparing the observed ToAs with the predicted ToAs given by Equation \ref{eq:phase}. It was not possible to obtain a white residual or a smooth timing residual for PSR~J1705$-$1906 for the whole data set, because the timing observations showed that PSR~J1705$-$1906 underwent two small glitches, as indicated in Figure \ref{fig:resiglt}. To investigate the spin behavior of PSR~J1705$-$1906, $\nu$ and $\dot\nu$ are obtained from independent fitting for short section of data, each of which typically spans $\sim$100 d  with overlapping of  $\sim$ 50 d.
Figure \ref{fig:glt} presents the variations of spin parameters after subtracting the pre-glitch timing model.
Two small glitches are visible which occurred with jumps in spin frequency of $\Delta\nu \sim 0.9\times10^{-9}$\,Hz and $ \sim 1.0\times10^{-9}$\ Hz, respectively. Comparing with the spin-frequency in post-glitches for PSR~J1705$-$1906, the observations do not appear to have any relaxation.
\citet{yu2013} also observed similar phenomenon in PSR~J0834$-$4159, PSR~J1413$-$6164, PSR~J1801$-$2304. In our case, the jump parameters obtained from the observations are likely permanent or long-term values.

If the gap between observations around a glitch is not too large, the glitch epoch can be estimated more accurately by requiring a phase-connected solution over the gap in the $\textsc{tempo2}$ fit.
However, if the gap is too large,
the glitch epoch is estimated from the ToA at the mid-point of the last pre-glitch and the first post-glitch, with the quoted uncertainty covers the gap in the data. For PSR~1705$-$1906, as the gap between observations around the glitch is too large, the estimated epochs for the two glitches are MJD~55199(6) (2010 01 03) and MJD~55953(16) (2012\,01\,27).

With the epochs fixed, the pre- and the post-glitch spin parameters given by Table~\ref{tbl:slt} can be determined. We extrapolate the spin parameters to the glitch epochs and calculate the fraction jump in $\nu$ and $\dot\nu$. The results are given in the third and fourth columns in Table~\ref{tab:glitch}.
The fifth and sixth columns give the results for the glitch parameters from directly fitting in $\textsc{tempo2}$. The values of glitch parameters are $\Delta\nu_{\rm{g}}/\nu=0.29(5)\times10^{-9}$ and $\Delta\dot{\nu}_{\rm{g}}/\dot{\nu}=0.3(5)\times10^{-3}$ around MJD~55199, and $\Delta\nu_{\rm{g}}/\nu=0.27(1)\times10^{-9}$ and $\Delta\dot{\nu}_{\rm{g}}/\dot{\nu}=0.5(2)\times10^{-3}$ around MJD~55953, respectively.
 We find that the parameter values and their corresponding uncertainties obtained from extrapolation and fitting {\bf are  consistent.}

\section{pulse profiles of PSR~J1705$-$1906}\label{s:pulse profiles}
\begin{figure}
\centerline{\includegraphics[angle=0,width=0.45\textwidth]{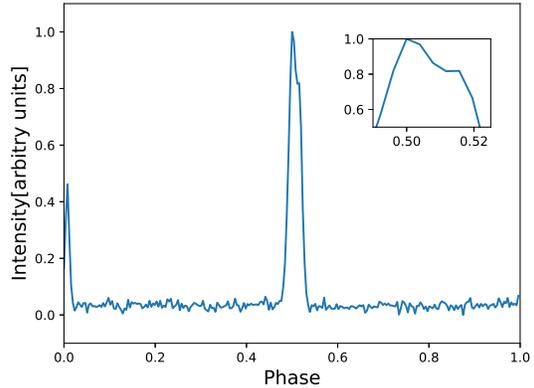}}
\vspace{-2mm}
\caption{The pulse profile of PSR~1705$-$1906  at 1540 MHz obtained {\bf using} Nanshan 25-m radio telescope.}
\label{fig:2pls}
\end{figure}

Figure \ref{fig:2pls} shows the pulse profile of PSR~J1705$-$1906 obtained at 1540\,MHz using the Nanshan 25-m radio telescope. The pulse profile consists of a main pulse and an interpulse, with separation of $\sim$0.5 phase from each other.
The main pulse contains two components with the intensity in the trailing component lower than that of the leading component by about {\bf 20\%}. The $W_{50}$ and $W_{10}$ are given by 0.009\,s and 0.014\,s, which correspond to 3\% and 4.7\% of the pulse period, respectively, where $W_{50}$ and $W_{10}$ are the full width at the $50\%$ and $10\%$ levels of pulse peak, respectively.  The interpulse has a fraction intensity of $\sim$0.43, weaker than the half-peak intensity of the main pulse. It is obvious from Figure \ref{fig:2pls} that the width of the interpulse is less than that of the main pulse.

\begin{figure}
\centerline{\includegraphics[angle=0,width=0.45\textwidth]{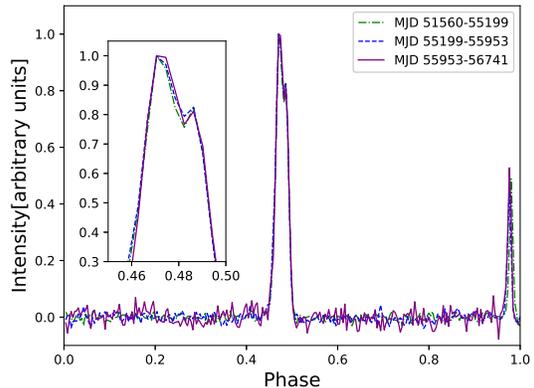}}
\vspace{-2mm}
\caption{The pulse profiles of PSR~1705$-$1906 from three different time segments obtained using Nanshan 25-m radio telescope at 1540 MHz.}
\label{fig:pulse}
\end{figure}

 Changes in pulse profile in association with glitches have been reported for several pulsars
 such as PSR J0742$-$2822 \citep{ksj13}, 1119$-$6227 \citep{wel11} and 2037+3621 \citep{kou18}.
In order to investigate whether such a link  also exists in PSR J1705$-$1906, we carefully examined the average pulse profile in the pre- and post-glitch 
data collected from 
Nanshan at 1540\,MHz. Figure~\ref{fig:pulse} shows the normalized average pulse profile from three different time segments with the pulse phase aligned. We found that the pulse width and the separation between main pulse and interpulse remain unchange between pre- and post-glitch.

\begin{table*}
\caption{The glitch size of pulsars with similar characteristic age and spin parameters to PSR~J1705$-$1906.}
\label{tab:otherglitch}
\begin{center}
\begin{tabular}{lllllll}
\toprule[1px]
PSR   &   P          & $\rm{\dot{P}}$ & $\tau_{\rm c} $ &     Epoch & $\Delta\nu_{\rm{g}}/\nu$&References \\
       &(s) &($\times10^{-15}$ s$^{-1}$)& ($\times10^{6}$ yr) &(MJD)    &($\times10^{-9}$)&\\
\hline
 J0611+1436 & 0.27033 &  3.997  & 1.07 & 55818  & 5575.3 &\citet{lyne2017}\\
 J1705$-$1906 & 0.29899 & 4.1379  & 1.14 & 48902 & 0.4(1) &\citet{espinoza2011} \\
        &&&      & 55199 & 0.29(5) & this work \\
          &&&    & 55953 & 0.27(1) & this work \\
J1812$-$1718 & 1.20537  &19.077 & 1.0 &49932(1) &1.5(2)&\citet{espinoza2011}  \\
    &&& &53106.2(1) &14.8(2)&\citet{espinoza2011}  \\
    &&& &54365.8(3)&1.4(1)  &\citet{espinoza2011}  \\
    J1902+0615 &0.67350 & 7.713 & 1.38 &48653.7(1)&0.4(1)&\citet{espinoza2011} \\
    &&& &49447(1)&0.3(1)&\citet{espinoza2011} \\
    &&& &50316(2)&0.3(1)&\citet{espinoza2011} \\
    &&& &51136(4)&0.4(1)&\citet{espinoza2011} \\
    &&& &54239(1)&0.26(3)&\citet{espinoza2011} \\
J2225+6535 & 0.68254 & 9.661& 1.12&43072(40)&1707(1) &\citet{backus1982} \\
   && & &51900&0.14(3)&\citet{janssen2006} \\
   && & &52950&0.08(4)&\citet{janssen2006} \\
   && & &53434(13)&0.2(1)&\citet{janssen2006} \\
   && & &54266(14)&0.36(8)&\citet{janssen2006} \\
J2257+5909  & 0.36824  & 5.753   &  1.01 & 49488.2(2) &0.75(4)&\citet{espinoza2011} \\
\hline
J1818$-$1422 & 0.29149 & 2.036 & 2.27 & 52057(7) &0.54(5) &\citet{yuan2010}\\
 J1957+2831 &0.30768 & 3.11 &1.57 & 52485(3) &0.3(1) &\citet{espinoza2011}\\
& && &52912(3)&0.13(3)&\citet{espinoza2011}\\
& && &54692.8(3)&5.8(3) &\citet{espinoza2011}\\
\toprule[1px]
    \end{tabular}
  \end{center}
\end{table*}

\section{Discussion and conclusions}\label{s:Discussion and conclusions}
In this paper we have
\begin{enumerate}
  \item updated the position, proper motion, distance and velocity of PSR~J1705$-$1906 in the equatorial  coordinates;
  \item calculated the  $\Delta_8$ value, which allowed the assessment of the size of the timing noise, and obtained a better timing red noise model using Cholesky solution;
  \item identified two new glitches and obtained their parameters.
\end{enumerate}

The new two glitches from PSR~J1705$-$1906 are quite small and do not appear to have significant relaxation, and have amplitudes similar to the 1992 event.
In our case, the jumps obtained from the observations are permanent or long-term values.
If we assume the pulsar glitched three times since it was discovered, then using the glitch activity parameter, which is defined as
\begin{equation}
A_{g}=\frac{1}{T}\sum\frac{\Delta\nu_{\rm{g}}}{\nu},
\end{equation}
where $\emph{T}$ is the total data span \citep{mckenna1990}, a value of $\rm A_{\rm g}\sim 2.0\times10^{-11}$ yr$^{-1}$ is obtained.
This is consistent with the assumption that relatively old pulsars have low glitch activity ($A_{\rm g}$) due to small and rare glitches \citep{espinoza2011,hobbs2010,lyne2000,yuan2010}.
A correlation is reported between the mean glitch rate and spin down rate with $ \langle\dot{N}_g\rangle \varpropto|\dot \nu|^{0.47(4)}$\citep{espinoza2011}  suggesting that  pulsars with low spin-down rates tend to exhibit small glitches.
 It is noted that this correlation is based on a large sample, which includes those pulsars that are yet
 to be detected with glitches.
For PSR J1705$-$1906 with $|\dot \nu|\sim46.288\times10^{-15}\,s^{-2}$,
 the predicted  $ \langle\dot{N}_g\rangle $ is
$\sim$ 0.019(2) yr$^{-1}$.
  The observed  mean glitch rate is large with a value of $\sim$0.083 yr$^{-1}$, which
is about 4.4 times that of the predicted value. This suggests that, in term of glitch, PSR J1705$-$1906 is
more active that those pulsars with similar spin down rate.

  For radio pulsars with characteristic age in the range between $1.0\times10^{6}$ and $1.4\times10^{6}$ yr, a total of 18
  glitches\footnote{\bf\url{http://www.atnf.csiro.au/people/pulsar/psrcat/glitchTbl.html}} were identified in six pulsars including PSR J1705$-$1906, of which four pulsars (PSRs J1705$-$1906, J1812$-$1718, J1902+0615, J2225+6535) have each at least three glitches detected {\bf \citep{backus1982,janssen2006,espinoza2011}}. For those radio pulsars with similar period (from 0.29 s to 0.31 s) and  similar $\dot{P}$ in the range between $2.0\times10^{-15}$ and $6\times10^{-15}$ s s$^{-1}$, at least four glitches were identified in two pulsars, PSRs~J1818$-$1422 and J1957+2831 \citep{yuan2010,espinoza2011}. The glitch sizes of these pulsars with similar characteristic age and spin parameters are given in Table~\ref{tab:otherglitch}. Almost all of these glitches have size of $\sim 10^{-9}$, except for the glitches in PSRs J0611+1436 and J2225+6535, which have sizes in the order of $\sim 10^{-6}$ {\bf \citep{backus1982, lyne2017}}. At this stage, we cannot exclude the possibility of occurrence of large glitch in PSR J1705$-$1906,
 and so a continuous monitoring of the pulsar is worthwhile.

 Pulsar glitch is one of the very few instances through which we can study the interior of a neutron star and the
properties of matter at supernuclear density.
Glitches have been argued to be the result of either starquakes or from angular momentum exchange between the faster rotating interior superfluid and the solid crust \citep{haskell2015}.
Consider the starquake regime, we infer that the fractional change in the radius of this pulsar due to the two glitches is $\Delta r/r\sim 0.28(3)\times10^{-9}$ and the fractional change in moment of inertia is $\Delta I/I\sim -0.56(6)\times10^{-9}$.
In the angular momentum exchange model, the interior vortices migrate outward and transfer momentum to the crust causing the crust to spin up and resulting in a glitch.
\citet{eya2017} introduce the fractional moment of inertia (FMI), which estimates the fraction of neutron star components involved in the glitch process,
\begin{equation}
\frac{I_{res}}{I_c} = - \sum_{1}^{n} \frac{1}{\dot{\nu}_c} \frac{\Delta{\nu_i}}{t_i},
\end{equation}
where $I_{res}$ is moment of inertia of the momentum reservoir, $t_i$ is the time interval preceding the $i$th glitch, $\dot{\nu}_c$ is the spin frequency derivatives of the crust and $I_c$ is the moment of inertia of the solid crust and all other components (superfluid interior and/or core) that strongly coupled to it.
The FMI  for 26 frequently glitching pulsars ranges from  $1.0\times 10^{-6}$ to  1.8 with a mean value of $10^{-3}$ and a distribution of the FMI that resembles a normal distribution with peak at $10^{-2}$.  We obtained  $I_{res}/I\approx 3.37\times10^{-4}$ for PSR~J1705$-$1906, which is consistent with the correlation between the fraction glitch size and the fractional moment of inertia \citep{eya2017}.

\section*{Acknowledgements}\label{s:Acknowledgements}

This work is supported by West light Foundation of CAS, Grant No. XBBS201421; Strategic Priority Research Programme of Chinese Academy of Sciences, Grant No. XDB23010200;  and National Natural Science Foundation of China (NSFC No. U1531137, 11373011). The FAST Fellowship is supported by Special Funding for Advanced Users, budgeted and administrated by Center for Astronomical Mega-Science, Chinese Academy of Sciences (CAMS).
JBW acknowledges support from NSFC, Grant 11403086, U1431107, 11573059). NW is supported by National Basic Research Program of China (2015CB857100), the National Program on Key Research and Development Project, Grant No. 2016YFA0400804. RY acknowledges supports from NSFC Project no. 11573059, and the West Light Foundation of the Chinese Academy of Sciences, project 2016-QNXZ-B-24.

This work is based on observations made with the Urumqi Nanshan 25 m Telescope,
which is operated by XAO and the Key Laboratory of Radio Astronomy, Chinese
Academy of Sciences. The Parkes radio telescope is part of the Australia Telescope,
which is funded by the Commonwealth of Australia for operation as a National
Facility managed by the Commonwealth Scientific and Industrial Research Organization.

%



%

\end{document}